\documentstyle[aps]{revtex}
\begin{document}
\title{Comment on the Roberts solution for the spherically-symmetric 
Einstein-scalar field equations} 
\author{Lior M. Burko}
\address{
Department of Physics, Technion---Israel institute of 
technology, 32000 Haifa, Israel}
\date{August 26, 1996}
\maketitle

\begin{abstract}
We critically examine the Roberts homothetic solution for  
the spherically symmetric Einstein-scalar field equations in double 
null coordinates, and show that the Roberts solution indeed solves 
the field equations only for one non-trivial case. We generalize this 
solution and discuss its relations with other known exact solutions.
\end{abstract}
Recently, there has been growing interest in exact solutions to the 
spherically symmetric Einstein-scalar field equations. This interest 
is motivated by Choptuik's discovery of critical phenomena 
in gravitational collapse \cite{choptuik}---and by efforts to understand 
this critical behaviour both analytically \cite{brady,oshiro,husain} 
and numerically \cite{stewart}---and by attempts to find counterexamples 
to the Cosmic Censorship Hypothesis \cite{roberts}. 

The homothetic exact solution by Roberts \cite{roberts} has captured 
lately much interest \cite{brady,husain,de oliviera}. Of the various 
solutions which Roberts discusses in Ref. \cite{roberts}, we shall focus 
in what follows on the homothetic solution in double null coordinates. 
(See also the solution given by Sussman \cite{sussman}.)  
[These coordinates deserve special attention, as it 
turns out that double null coordinates are especially 
convenient for the study of 
critical phenomena, as one can derive numerically stable 
codes for near-critical evolutions using double null coordinates. 
(See discussion in Ref. \cite{stewart}).] 

Following Roberts \cite{roberts} we write the line element for the 
spherically-symmetric spacetime as
\begin{equation}
\,ds^{2}=-\,du\,dv+r^{2}(u,v)\,d\Omega^{2},
\label{metric}
\end{equation}
where $\,d\Omega^{2}$ is the regular line element of the unit sphere. 
The Einstein-scalar field equations reduce then to the dynamical equations
\begin{equation}
\Phi_{,uv}+\left(r_{,u}\Phi_{,v}+r_{,v}\Phi_{,u}\right)/r=0
\label{scalar}
\end{equation}
\begin{equation}
r_{,uv}+\left(r_{,u}r_{,v}+\frac{1}{4}\right)/r=0
\label{E1}
\end{equation}
\begin{equation}
\left(2r_{,u}r_{,v}+\frac{1}{2}\right)/r^{2}-2\Phi_{,u}\Phi_{,v}=0
\label{E2}
\end{equation}
and to the constraint equations
\begin{equation} 
r_{,vv}+r\left(\Phi_{,v}\right)^{2}=0
\label{E3}
\end{equation}
\begin{equation}
r_{,uu}+r\left(\Phi_{,u}\right)^{2}=0,
\label{E4}
\end{equation}
where $\Phi$ is a real scalar field. 

The solution by Roberts (Eqs. (42) of Ref. \cite{roberts}) reads 
\begin{eqnarray}
r^{2}&=&
\frac{1}{4}(1+2\sigma)[v-(1+2\sigma)u][(1-2\sigma)v-(1+2\sigma)^{2}u]
\nonumber \\
\Phi&=&\frac{1}{2}\ln\left | 1-\frac{4\sigma}{(1+2\sigma)\left[1-
\frac{u(1+2\sigma)}{v}\right]}\right | .
\label{sol}
\end{eqnarray}
Allowing $\sigma$ to be an arbitrary function of $v$, one finds that 
with Eqs. (\ref{sol}), the integrability conditions for  
Eqs. (\ref{scalar})--(\ref{E3}) are, corrspondingly, 
\begin{equation}
\sigma\sigma ' v=0
\label{in1}
\end{equation}
\begin{equation}
\sigma '\left(2A-2\sigma v -v\right)-\sigma(1+\sigma )=0
\label{in2}
\end{equation}
\begin{equation}
\sigma 'A-
\sigma(1+\sigma )=0
\label{in3}
\end{equation}
\begin{equation}
\sigma ''\left(Au-2\sigma uv-uv-\sigma v^{2}\right)
+2\sigma '^{2}u\left(4\sigma^{2}u
+4\sigma +u-v\right)-2\sigma '\sigma v=0,
\label{in4}
\end{equation}
where $A=8\sigma^{3}u+12\sigma^{2}u+6\sigma u+u$. Here, a prime denotes 
partial differentiation with respect to $v$. [Eq. (\ref{E4}) is satisfied 
for any $\sigma (v)$.] 

In order that (\ref{sol}) should be a solution of the Einstein-scalar field 
equations, the entire set of equations (\ref{in1})--(\ref{in4}) should 
be satisfied. It can be readily verified, that the only solutions of Eqs. 
(\ref{in1})--(\ref{in4}) are $\sigma = 0$ or $\sigma=-1$. 
The former is nothing but the 
trivial vacuum solution (i.e., Minkowski spacetime), 
while the latter indeed represents a solution, 
which is, however, just a particular case of the solution given by 
Refs. \cite{brady,oshiro,sussman}. (This case 
is equivalent to the solution of Ref. \cite{oshiro} when the  
parameter $p$ of Ref. \cite{oshiro} is set equal to $-2$.) Any other 
$\sigma (v)$, and in particular any other constant $\sigma$, does not solve 
the Einstein-scalar field equations.

Consequently, the solution by Roberts \cite{roberts} may represent just 
a class of measure zero among the solutions to the spherically-symmetric 
homothetic Einstein-scalar field equations given by 
Refs. \cite{brady,oshiro,sussman}. However, we can understand the 
relation between the solution of Ref. \cite{roberts} and the solutions 
of Refs. \cite{brady,oshiro,sussman} by transforming the solution given by  
Eq. (28) of Ref. \cite{roberts} to double null coordinates. Defining 
\begin{equation}
u=(1+2\sigma)v-2r
\label{trans}
\end{equation}
(note the difference between this definition and 
Eq. (38) of Ref. \cite{roberts}), one readily finds that the correct 
expression for the solution in double null coordinates is given by
\begin{equation}
r^2=\frac{1}{4}\left[\left(1-4\sigma ^2\right)v^2-2uv+u^2\right]
\label{sol1}
\end{equation}
\begin{equation}
\Phi=\pm\frac{1}{2}\ln \left | 1-\frac{4\sigma}{1+2\sigma-u/v} \right |.
\label{sol2}
\end{equation}
One immediately notices that this is nothing but the solution given by 
Oshiro {\em et al} \cite{oshiro}, with the parameter $p$ of Ref. 
\cite{oshiro} satisfying $p=2\sigma$. Now, we can understand why the 
generally incorrect solution by Roberts gives a correct result 
for $\sigma=0,-1$: For these particular cases, our coordinate transformation 
(\ref{trans}) coincides with the transformation given by Eq. (38) of Ref. 
\cite{roberts}. However, the latter does not preserve the form of the 
metric (\ref{metric}), while transformation (\ref{trans}) does. 
(In fact, the coordinate transformation of Ref. \cite{roberts} yields 
$g_{uv}=-\frac{1}{2}(1+2\sigma)^{2}$, instead of the $-\frac{1}{2}$ value 
implied by Eq. (\ref{metric}). One can see again that for $\sigma=0,-1$ 
the Roberts solution coincides with the solution of Refs. 
\cite{brady,oshiro,sussman}.)

Now, we are in a position to try to generalize the solution: 
Recall, that in 
Ref. \cite{oshiro} the parameter $p$ is assumed to be constant. Allowing 
$\sigma$ to be an arbitrary function of $v$, we readily find that the 
integrability condition of Eqs. (\ref{scalar})--(\ref{E3}) is 
\begin{equation}
\left(\sigma ''v+2\sigma '\right)\sigma v=0.
\label{integ}
\end{equation}
[This integrability condition arrises from the constraint equation 
(\ref{E3}). The other Einstein-scalar field equations are satisfied 
identically for any $\sigma (v)$.] 
Eq. (\ref{integ}) is integrated to $\sigma (v)=a+b/v$, 
with $a$ and $b$ being constants of integration. Apparently, we now have 
a generalized solution which depends on two parameters, namely, on $a$ and 
$b$. However, transforming to the new coordinates $\bar{v}=v+b/a$ and 
$\bar{u}=u+b/a$ we find that the solution is again given by Eqs. 
(\ref{sol1}) and (\ref{sol2}), this time with the newly-defined coordinates 
$\bar{u},\bar{v}$ and with the {\em constant} parameter $a$. 
[This coordinate transformation does not change the metric form 
(\ref{metric}).]  
Namely, we have shown that this solution is the most general within this 
class of solutions. In particular, 
the assumption of Ref. \cite{oshiro} that the solution is given by 
a constant $\sigma$ is unnecessary, as we have shown that the constancy 
of $\sigma$  can always be achieved for an appropriate choice of the 
coordinates. [Consequently, the most general solution depends on just one 
free parameter (see Ref. \cite{de oliviera}, where the solution is given in 
terms of two parameters).]  
However, it might be usefull to express the solution in  
coordinates other than $\bar{u},\bar{v}$, e.g., 
in the testing of numerical codes. 

I thank Patrick Brady, Amos Ori and Mark Roberts for useful discussions.


\begin{thebibliography}{99}

\bibitem{choptuik} Choptuik M.W. (1993) {\em Phys. Rev. Lett.} {\bf 70} 9.

\bibitem{brady} Brady P.R. (1994) {\em Class. Quantum Grav.} {\bf 11} 1255.

\bibitem{oshiro} Oshiro Y., Nakamura K., and Tominatsu A. (1994) 
{\em Prog. Theor. Phys.} {\bf 91} 1265.

\bibitem{husain} Husain V., Martinez E.A., and N\'{u}\~{n}ez D. (1994), 
{\em Phys. Rev. D} {\bf 50} 3783.

\bibitem{stewart} Hamad\'{e} R.S. and Stewart J.M. (1996) {\em Class. 
Quantum Grav.} {\bf 13} 497. 

\bibitem{roberts} Roberts M.D. (1989) {\em Gen. Relativ. Gravitation} 
{\bf 21} 907.

\bibitem{de oliviera} de Oliveira H.P. and Cheb--Terrab E.S. (1996) 
{\em Class. Quantum Grav.} {\bf 13} 425.

\bibitem{sussman} Sussman R.A. (1991) {\em J. Math. Phys.} {\bf 32} 223.

\end{thebibliography}
\end{document}